\documentclass[pre,aps,twocolumn,superscriptaddress]{revtex4-1}

\usepackage{graphicx}
\usepackage{amssymb,amsfonts,amsmath}
\usepackage{color}
\usepackage{ulem}
\usepackage[hidelinks]{hyperref}
\usepackage{bbm}
\usepackage{bbold}
\usepackage[scr=rsfs,cal=boondox]{mathalfa}

\usepackage{tikz}
\usetikzlibrary{shapes}
\usetikzlibrary{patterns}
\usetikzlibrary{angles, quotes}

\newcommand{\rv}{{\mathbf r}}

\newcommand{\Tr}{{\rm Tr}\,}
\newcommand{\e}{{\rm e}}

\newcommand{\pv}{{\bf p}}

\newcommand{\ms}{\color{black}}

\newcommand{\msphantom}[1]{$\ldots$}

\newcommand{\eqr}[1]{Eq.~\eqref{#1}}

\newcommand{\mydelete}[1]{{}}

\newcommand{\rmexc}{{\rm exc}}
\newcommand{\rmext}{{\rm ext}}

\newcommand{\rmid}{{\rm id}}

\newcommand{\sumi}{\sum_{i{\ms =1}}^{{\ms N}}}
\newcommand{\Ham}{H_{\ms N}}

\begin{document}

\title{\ms Entropy functional theory for inhomogeneous fluids}

\author{Matthias Schmidt}
\affiliation{Theoretische Physik II, Physikalisches Institut, 
  Universit{\"a}t Bayreuth, D-95447 Bayreuth, Germany}
\email{Matthias.Schmidt@uni-bayreuth.de}

\date{26 June 2026, \ms revised: 21 October 2026}

\begin{abstract}
  We present an exact variational scheme for the physics of
  inhomogeneous classical fluids in thermal equilibrium. A joint
  metadensity minimization principle is proven for the one-body
  density and the global interparticle distance distribution.  The
  theory bypasses the inhomogeneous two-body density and thus remains
  computationally simple. A universal excess entropy functional
  accounts for all many-body correlations in arbitrary pairwise
  interacting systems.  The framework is relevant for neural
  functional machine learning, for soft matter design, {\ms and for
    predicting hyperobservables,} structural correlation functions,
  {\ms and first-principles mechanisms of entropy-energy competition}.
\end{abstract}

\maketitle

Classical density functional theory \cite{hansen2013, evans1979,
  evans1992, evans2016} is a powerful approach for the description of
the equilibrium properties of liquids and more general soft matter
systems. The framework is formally exact and typically it requires one
to choose approximations for the central density functional
relationships. Much inspiration stems from the rich heritage of
analytical, closed-form approximations, including Rosenfeld's
fundamental-measure theory for hard sphere mixtures
\cite{rosenfeld1989, roth2010}, the celebrated mean-field
approximation for penetrable soft core
behaviour~\cite{archer2017meanField}, as well as more sophisticated
weighted-density approximation schemes \cite{denton1989, denton1990,
  denton1991, belloni2026}.  These theories account in their
respective ways for the system-specific nonlocal correlation structure
between the particles.

Recent progress in neural functional learning has demonstrated that
methods of supervised machine learning, based on many-body simulation
data, constitute an excellent means to overcome the limitations of the
analytically known density functional approximations~\cite{cats2022ml,
  kelley2024ml, yatsyshin2022, malpica-morales2023, monti2026,
  fang2022, yang2024, pan2025, dijkman2024ml, ram2025ddft,
  teixera2014, lin2019ml, lin2020ml, glitsch2025disks,
  simon2024patchy, simon2023mlPatchy, bui2024neuralrpm,
  bui2025dielectrocapillarity, zhou2026azeoptropic, bui2026abinitio}.
In particular the local learning strategy by Samm\"uller~{\it
  et~al.}~\cite{sammueller2023neural, sammueller2023whyNeural,
  sammueller2024attraction, robitschko2025mixShort,
  sammueller2024hyperDFT, sammueller2024whyhyperDFT,
  sammueller2024pairmatching, kampa2024meta, kampa2026pairmatching,
  kampa2026spherical, sammueller2025chemicalPotential,
  delasheras2023perspective, zimmermann2024ml} proves to be highly
flexible and adaptable to extended frameworks, such as the
hyperdensity functional theory for general observables
\cite{sammueller2024hyperDFT, sammueller2024whyhyperDFT}, the
metadensity functional theory for general pair interaction potentials
\cite{kampa2024meta, kampa2026pairmatching, kampa2026spherical}, the
power functional theory for nonequilibrium dynamics
\cite{schmidt2022rmp, delasheras2023perspective, zimmermann2024ml},
and the very recent formulation for the {\it ab initio} description of
liquids~\cite{bui2026abinitio}. The local learning approach was
partnered with and contrasted against Percus' exact density functional
for one-dimensional hard rods, see the introductory
Ref.~\cite{sammueller2023whyNeural}.

Entropy is a founding principle for the thermal description of matter
\cite{hansen2013} and it features prominently in liquid state theory,
see e.g.~Refs.~\cite{dedominicis1964, raveche1971, wallace1987,
  baranyai1989, laird1992, huang2024}.  Percus discussed entropy in
several of his seminal classical density functional contributions
\cite{percus1996, percus1994aspects, percus1989entropy}.  Relevant
recent work was carried out by Frusawa \cite{frusawa2018, frusawa2019,
  frusawa2021}, Keffer and coworkers \cite{nicholson2021, sluss2022},
and Shirai {\it et~al.}~\cite{shirai2025}.  A local entropy density
was introduced~\cite{schmidt2011internalEnergy} and the closely
related local thermal susceptibility was demonstrated to constitute a
local measure of entropic fluctuations \cite{eckert2020,
  eckert2023fluctuation, coe2022pre}.  The physics that is involved
was epitomized succinctly by Chandler who noted that ``liquids are
strongly interacting but disordered systems whose very existence
implies a delicate balance between energy and
entropy''~\cite{chandler2017}.

The recent metadensity functional approach by Kampa {\it
  et~al.}~\cite{kampa2024meta, kampa2026pairmatching,
  kampa2026spherical} offers the prospect to work efficiently with
general forms of the pair potential $\phi(r)$ that characterizes the
fluid. Although the functional dependence on~$\phi(r)$ lies at the
heart of density functional theory \cite{evans1979, evans1992}, this
fact has thus far found very little practical implementation. The
neural functional accessibility \cite{kampa2024meta,
  kampa2026pairmatching, kampa2026spherical} of the functional
dependence on $\phi(r)$ is important for problems of inverse soft
matter design, such as implementing Henderson's theorem
\cite{henderson1974uniqueness} to retrieve the pair potential from the
given bulk fluid pair structure \cite{kampa2024meta,
  kampa2026spherical}, as well as for the targeted design of
inhomogeneous states \cite{kampa2024meta}.  The functional dependence
on $\phi(r)$ allows one to access the global distribution
(`histogram') of interparticle distances~$G(r)$ in inhomogeneous
systems and also to use exact functional relationships
\cite{evans1992} as a systematic basis for regularized functional
learning \cite{kampa2026pairmatching}.

Here we provide an alternative route towards the description of soft
matter that challenges conventional density functional wisdom. An
entropic metadensity functional is identified based on the standard
one-body density $\rho(\rv)$ and the global distribution of
interparticle distances $G(r)$. The functional dependence on $G(r)$
allows one to extract explicitly the pair potential as a mere linear
contribution to the overall free energy. The nontrivial dependencies
on $\rho(\rv)$ and $G(r)$ are encapsulated in a universal excess (over
ideal gas) entropy functional, which applies across all Hamiltonians
with pairwise interparticle potential energy.  The approach avoids
having to engage with the inhomogeneous two-body density distribution
\cite{percus1996, percus1994aspects, percus1989entropy}, yet it
retains both the full correlation structure of the many-body system as
well as the computational simplicity of standard density functional
theory.

We consider Hamiltonians that possess the following standard form:
\begin{align}
  \Ham &= \sumi \frac{\pv_i^2}{2m} + u(\rv^N) + \sumi V_\rmext(\rv_i),
  \label{EQentropyHamiltonian}
\end{align}
with mass $m$, position $\rv_i$ and momentum $\pv_i$ of particle $i$,
{\ms and number of (identical) particles $N$}.  The interparticle
potential $u(\rv^N)$ depends on all particle position coordinates
$\rv^N= \rv_1,\ldots, \rv_N$ and $V_\rmext(\rv)$ is an external
potential, which is here expressed as a function of a generic position
variable~$\rv$.  We take the interparticle potential energy~$u(\rv^N)$
to consist of only pairwise contributions,
\begin{align}
  u(\rv^N) &= \frac{1}{2}\sum_{i=1}^N \sum_{j=1,j\neq i}^N\phi(|\rv_i-\rv_j|)
  \label{EQentropyInterparticlePotentialPairwise}  \\
  &=  \int_0^\infty dr \hat G(r) \phi(r).
  \label{EQentropyInterparticlePotentialReWriting}
\end{align}
The re-writing \eqref{EQentropyInterparticlePotentialReWriting} uses
the characteristic function $\hat G(r)$, which constitutes an
`instantaneous' histogram of interparticle distances $r$ in the
system~\cite{kampa2024meta, kampa2026pairmatching,
  kampa2026spherical},
\begin{align}
  \hat G(r) &=
  \frac{1}{2}\sum_{i=1}^N \sum_{j=1,j\neq i}^N \delta(r-|\rv_i-\rv_j|),
  \label{EQentropyGhat}
\end{align}
where $\delta(\cdot)$ denotes the Dirac distribution.  Using the
standard density operator $\hat \rho(\rv) = \sumi\delta(\rv-\rv_i)$
allows one to express similarly the external potential energy [last
  term in \eqr{EQentropyHamiltonian}] as a position integral
\cite{evans1979, hansen2013},
\begin{align}
  \sumi V_\rmext(\rv_i) &= \int d\rv \hat\rho(\rv) V_\rmext(\rv).
  \label{EQentropyVextReWriting}
\end{align}
The integrals~\eqref{EQentropyInterparticlePotentialReWriting} and
\eqref{EQentropyVextReWriting} serve to formulate the Hamiltonian
\eqref{EQentropyHamiltonian} in the following alternative form:
\begin{align}
  \Ham = \sumi \frac{\pv_i^2}{2m} + \int_0^\infty dr \hat G(r)\phi(r)
  + \int d\rv \hat\rho(\rv) V_\rmext(\rv).
  \label{EQentropyHamiltonianRewritten}
\end{align}
One recovers via functional differentiation of
\eqr{EQentropyHamiltonianRewritten} the density operator,
$\hat\rho(\rv) = \delta \Ham/\delta V_\rmext(\rv)$, \cite{evans1979}
as well as the global distance distribution operator, $\hat G(r) =
\delta \Ham / \delta \phi(r)$ \cite{kampa2024meta,
  kampa2026pairmatching, kampa2026spherical}, see its
definition~\eqref{EQentropyGhat}.

We base the entropy metadensity functional construction on Mermin's
many-body functional \cite{evans1979, mermin1965, hansen2013},
\begin{align}
  \beta \Omega_M[f] &= \Tr f (\ln f + \beta \Ham - \beta \mu N),
  \label{EQentropyMerminFunctionalDefinition}
\end{align}
where the square brackets indicate the functional dependence on the
phase space probability distribution function $f(\rv^N,\pv^N)$. Here
$f$ constitutes a trial function that is non-negative and normalized,
$\Tr f=1$, but otherwise unspecified at this stage.  The standard
classical (grand canonical) trace is $\Tr\,\cdot\, = \sum_{N=0}^\infty
(h^{dN}N!)^{-1}\int d\rv^N d\pv^N\,\cdot\,$, where~$h$ indicates
Planck's constant, $d$ is the spatial dimensionality, and the phase
space integral is over the positions and momenta of all particles. The
thermodynamic parameters are the chemical potential~$\mu$ and the
absolute temperature $T$, with $\beta=1/(k_BT)$, where $k_B$ denotes
the Boltzmann constant.

One central property of Mermin's functional
\eqref{EQentropyMerminFunctionalDefinition} is to give the equilibrium
value of the grand potential $\Omega_0$ upon
minimization~\cite{mermin1965, evans1979, hansen2013},
\begin{align}
  \Omega_0  &= \min_f \Omega_M[f],
  \label{EQentropyMerminFunctionalMinimization}
\end{align}
where the minimization is performed over all (normalized) trial
functions $f(\rv^N,\pv^N)$.  We recall the grand potential in
elementary form, $\Omega_0=-k_BT\ln \Xi$, with grand partition sum
$\Xi=\Tr \e^{-\beta(\Ham-\mu N)}$. The minimizer of Mermin's
functional \eqref{EQentropyMerminFunctionalMinimization} is the
equilibrium grand distribution function, $f_0(\rv^N,\pv^N)=\e^{-\beta
  (\Ham-\mu N)}/\Xi$, such that $\Omega_0=\Omega_M[f_0]$
\cite{evans1979, mermin1965, hansen2013}.

To set up an extended classical constrained search we generalize
Refs.~\cite{levy1979, dwandaru2011} and introduce {\it two}
simultaneous constraints:
\begin{align}
  \rho(\rv) &= \Tr f \hat\rho(\rv),
  \label{EQentropyConstraint1}\\
  G(r) &= \Tr f \hat G(r),
  \label{EQentropyConstraint2}
\end{align}
where we recall $\hat\rho(\rv)$ as the one-body density operator and
$\hat G(r)$ as the global interparticle distance
operator~\eqref{EQentropyGhat}. The many-body distribution function
$f(\rv^N,\pv^N)$ constitutes a mere trial function at this stage and
both the resulting density profile $\rho(\rv)$ and the mean distance
histogram $G(r)$ inherit this (trial) status via the
constraints~\eqref{EQentropyConstraint1} and
\eqref{EQentropyConstraint2}.

We define the grand potential metadensity functional $\Omega[\rho,G]$
via the following constrained search:
\begin{align}
  \Omega[\rho,G] &= \min_{f\to\rho,G} \Omega_M[f],
  \label{EQentropyOmegaFunctionalFromMermin}
\end{align}
where the notation $f\to\rho,G$ indicates that the minimization is
performed in the reduced function space of only those $f(\rv^N,
\pv^N)$ that satisfy the two simultaneous
constraints~\eqref{EQentropyConstraint1} and
\eqref{EQentropyConstraint2}.  The value of the functional at the
constrained minimum then depends functionally on both $\rho(\rv)$ and
$G(r)$, as is indicated by $\Omega[\rho,G]$
in~\eqr{EQentropyOmegaFunctionalFromMermin}.

Per construction the {\it global} minimum is reached upon relaxing
both constraints and hence:
\begin{align}
  \Omega_0 = \min_{\rho,G} \Omega[\rho,G],
  \label{EQentropyMinimizationPrinciple}
\end{align}
where we recall $\Omega_0$ as the grand
potential~\eqref{EQentropyMerminFunctionalMinimization} for the
Hamiltonian \eqref{EQentropyHamiltonian}.  The benefit of the
constrained search is that the minimization principle
\eqref{EQentropyMinimizationPrinciple} now applies in the reduced
function space of $\rho(\rv)$ and $G(r)$, in contrast to the full
minimization~\eqref{EQentropyMerminFunctionalMinimization} of Mermin's
functional in the function space of $f(\rv^N, \pv^N)$.

At the global minimum \eqref{EQentropyMinimizationPrinciple} the
many-body distribution function attains its equilibrium value,
$f_0(\rv^N, \pv^N)$; we recall its Boltzmann form given below
\eqr{EQentropyMerminFunctionalMinimization}. Then the
constraints~\eqref{EQentropyConstraint1} and
\eqref{EQentropyConstraint2} represent the thermal equilibrium density
profile $\rho_0(\rv)$ and the global pair distance histogram
$G_0(r)$. These are given respectively by
\begin{align}
 \rho_0(\rv) &= \Tr f_0 \hat\rho(\rv)
 = \frac{\delta \Omega_0}{\delta V_\rmext(\rv)}\Big|_\phi,
 \label{EQentropyMinimizer1} \\
 G_0(r) &= \Tr f_0\hat G(r)
 =\frac{\delta \Omega_0}{\delta \phi(r)}\Big|_{V_\rmext}, 
 \label{EQentropyMinimizer2}
\end{align}
where the functional derivative expressions follow straightforwardly
from the functional chain rule; this requires one to identify
$\hat\rho(\rv)$ and $\hat G(r)$ as the specific functional derivatives
of~$\Ham$ given below the definition
\eqref{EQentropyHamiltonianRewritten}.

It is straightforward to evaluate the grand potential metadensity
functional~\eqref{EQentropyOmegaFunctionalFromMermin} at the joint
minimizers \eqref{EQentropyMinimizer1} and
\eqref{EQentropyMinimizer2}. The result is the equilibrium grand
potential,
\begin{align}
  \Omega_0 = \Omega[\rho_0,G_0],
\end{align}
and we drop the subscript of $\rho_0(\rv)$ and $G_0(r)$ in the
following for notational simplicity.  At the minimum, the functional
derivative with respect to each argument vanishes and hence {\it two}
Euler-Lagrange equations hold,
\begin{align}
  \frac{\delta \Omega[\rho,G]}{\delta\rho(\rv)}\Big|_G &= 0 \quad\text{(min)},
  \label{EQentropyEL1raw}  \\
  \frac{\delta \Omega[\rho,G]}{\delta G(r)}\Big|_\rho &=0 \quad\text{(min)}.
  \label{EQentropyEL2raw}
\end{align}

The additive form of the
Hamiltonian~\eqref{EQentropyHamiltonianRewritten} together with the
specific structure of Mermin's
functional~\eqref{EQentropyMerminFunctionalDefinition} allow one to
decompose $\Omega[\rho,G]$ by identifying $\rho(\rv)$ and $G(r)$ via
the constraints \eqref{EQentropyConstraint1} and
\eqref{EQentropyConstraint2}. This yields the sum
\begin{align}
  \Omega[\rho,G] &=  K[\rho]  - TS[\rho,G] 
  + \int_0^\infty dr G(r) \phi(r)
  \notag\\&\quad 
  +\int d\rv  \rho(\rv) [ V_\rmext(\rv) -  \mu],
  \label{EQentropyOmegaDecomposition}
\end{align}
where $K[\rho]=(d/2)k_BT\int d\rv\rho(\rv)$ is the standard classical
equilibrium mean kinetic energy functional and $S[\rho,G]$ is the
entropy functional, as specified below. The distance integral [third
  term in \eqr{EQentropyOmegaDecomposition}] constitutes the mean
interparticle potential energy as the thermal average of the
corresponding operator
identity~\eqref{EQentropyInterparticlePotentialReWriting}. The
position integral [fourth term in \eqr{EQentropyOmegaDecomposition}]
accounts for the external potential energy and the chemical potential
contribution, which both are standard \cite{evans1979, hansen2013}.

The ideal entropy density functional is known exactly,
\begin{align}
  S_\rmid[\rho] &= -k_B \int d\rv \rho(\rv)
  \big[\ln \big(\rho(\rv)\Lambda^d\big) -(d+2)/2 \big],
  \label{EQentropyNoninteractingGas}
\end{align}
where $\Lambda = h/\sqrt{2\pi m k_BT}$ denotes the thermal de Broglie
wavelength \cite{hansen2013} with Planck's constant $h$.  The
combination with the kinetic energy functional $K[\rho]$ yields the
intrinsic free energy density functional of the ideal gas, $
F_\rmid[\rho] = K[\rho] - TS_\rmid[\rho] = k_BT \int d\rv
\rho(\rv)\big[\ln(\rho(\rv)\Lambda^d)-1\big]$, which again is standard
\cite{evans1979, hansen2013}.

We have arrived at a stage where the nontrivial contribution to the
grand potential \eqref{EQentropyOmegaDecomposition} is the entropy
functional $S[\rho,G]$. Comparing the functional
decomposition~\eqref{EQentropyOmegaDecomposition} to the constrained
search \eqref{EQentropyOmegaFunctionalFromMermin} yields (upon
multiplication by $-1/T$) the following constrained {\it maximization}
form of the entropy functional:
\begin{align}
  & S[\rho,G] =  \max_{f\to \rho,G}
  \Tr f\Big( -k_B \ln f 
  -\frac{1}{T}\sumi\frac{\pv_i^2}{2m}
  \Big) + \frac{K[\rho]}{T}.
  \label{EQentropyViaConstrainedMaximization}
\end{align}
Inside of the constrained maximization
\eqref{EQentropyViaConstrainedMaximization} the scaled kinetic energy
acts similar to a Lagrange multiplier to generate the Maxwellian
velocity distribution. The resulting kinetic offset over the entropy
is removed again by the scaled kinetic energy functional $K[\rho]/T$
[last term in \eqr{EQentropyViaConstrainedMaximization}].

We split the total entropy functional
\eqref{EQentropyViaConstrainedMaximization} into ideal and excess
(over ideal gas) contributions,
\begin{align}
  S[\rho,G] &= S_\rmid[\rho] + S_\rmexc[\rho,G],
  \label{EQentropySidSexcSplitting}
\end{align}
where we recall the ideal gas form \eqref{EQentropyNoninteractingGas}.
The excess entropy functional $S_\rmexc[\rho,G]$ now contains all
equilibrium many-body correlation effects.  Per construction, for the
ideal gas $S_\rmexc[\rho,G_\rmid]=0$, where the ideal global distance
distribution is $ G_\rmid(r) = \int d\rv d\rv'\rho(\rv)\rho(\rv')
\delta(r-|\rv-\rv'|)/2$, as follows from the two-body density
factorizing trivially into $\rho(\rv)\rho(\rv')$.

Inserting the entropic splitting \eqref{EQentropySidSexcSplitting}
into the grand potential \eqref{EQentropyOmegaDecomposition} and
identifying the ideal gas free energy functional $F_\rmid[\rho]$
yields the decomposition
\begin{align}
  \Omega[\rho,G] &= 
   F_\rmid[\rho]  - TS_\rmexc[\rho,G] 
  +\int_0^\infty dr G(r) \phi(r)\notag\\
  &\quad
  +\int d\rv \rho(\rv)\big(V_\rmext(\rv)-\mu\big).
  \label{EQentropyOmegaDecompositionExcess}
\end{align}
The sum \eqref{EQentropyOmegaDecompositionExcess} consists of ideal
free energy, scaled excess entropy, interparticle potential energy,
and the joint external and chemical potential contributions.

To render the minimization principle
\eqref{EQentropyMinimizationPrinciple} more concrete, we define {\it
  entropic} direct correlation functionals via the following
functional derivatives:
\begin{align}
  c_\rho(\rv;[\rho,G])  &= 
  \frac{\delta S_\rmexc[\rho,G]/k_B}{\delta\rho(\rv)}\Big|_G,
  \label{EQentropyDirectEntropic1}\\
  c_G(r;[\rho,G]) &= 
  \frac{\delta S_\rmexc[\rho,G]/k_B}{\delta G(r)}\Big|_\rho.
  \label{EQentropyDirectEntropic2}
\end{align}
Inserting the decomposition \eqref{EQentropyOmegaDecompositionExcess}
into the Euler-Lagrange equations \eqref{EQentropyEL1raw} and
\eqref{EQentropyEL2raw} allows one to re-write these, upon carrying
out the functional derivatives and identifying the direct correlation
functionals \eqref{EQentropyDirectEntropic1} and
\eqref{EQentropyDirectEntropic2}, as:
\begin{align}
  c_\rho(\rv;[\rho,G])  &= 
  \ln(\rho(\rv)\Lambda^d) + \beta V_\rmext(\rv) - \beta\mu,
  \label{EQentropyEL1rhoG}\\
  c_G(r;[\rho,G]) &= \beta\phi(r).
  \label{EQentropyEL2rhoG}
\end{align}
Notably, the `position' Euler-Lagrange
equation~\eqref{EQentropyEL1rhoG} shares its right hand side with the
standard Euler-Lagrange equation of classical density functional
theory~\cite{hansen2013, evans1979}.  The `distance' Euler-Lagrange
equation~\eqref{EQentropyEL2rhoG} has thus far not been identified in
the literature, to the best of our knowledge, and it carries an
equally fundamental status as \eqr{EQentropyEL1rhoG}.

The standard splitting of the grand potential density functional
$\Omega[\rho]$ consists of the intrinsic free energy functional
$F[\rho]$ and the external and chemical contributions,
$\Omega[\rho]=F[\rho]+\int d\rv\rho(\rv)[V_\rmext(\rv)-\mu]$. In
contrast, the present approach yields, from re-organizing the
decomposition \eqref{EQentropyOmegaDecomposition}, the following
entropic functional version of the standard thermodynamic sum:
\begin{align}
  \Omega[\rho,G] &=
  E[\rho,G]  -TS[\rho,G]  - \mu N[\rho],
  \label{EQmetaentropyOmegaViaEnergyEntropy}
\end{align}
where $E[\rho,G]$ denotes the mean total energy (specified below), the
entropy functional $S[\rho,G]$ is defined via the many-body
search~\eqref{EQentropyViaConstrainedMaximization} and we recall its
ideal-excess splitting \eqref{EQentropySidSexcSplitting}, and $N[\rho]
= \int d\rv \rho(\rv)$ is the trivial mean total particle number
functional.  The mean energy functional $E[\rho,G]$ is the
straightforward sum of the kinetic and the interparticle and external
potential energy contributions,
\begin{align}
  E[\rho,G] &= K[\rho] 
  + \int_0^\infty dr G(r) \phi(r)
  + \int d\rv \rho(\rv) V_\rmext(\rv).
  \label{EQmeanEnergyFunctional}
\end{align}
Notably the pair potential $\phi(r)$ remains isolated solely in the
radial integral [second term in \eqr{EQmeanEnergyFunctional}]. All
further contributions in both Eqs.~\eqref{EQmeanEnergyFunctional} and
\eqref{EQmetaentropyOmegaViaEnergyEntropy} are independent of
$\phi(r)$. Hence the excess entropy functional $S_\rmexc[\rho,G]$, as
it is defined via Eqs.~\eqref{EQentropyViaConstrainedMaximization} and
\eqref{EQentropySidSexcSplitting}, contains all nontrivial
interparticle correlation effects. All further functionals are known
explicitly. It is the joint minimization principle
\eqref{EQentropyMinimizationPrinciple}, or equivalently the pair of
Euler-Lagrange equations \eqref{EQentropyEL1rhoG} and
\eqref{EQentropyEL2rhoG}, that determine $\rho(\rv)$ and $G(r)$ in
equilibrium for any given system. Evaluating the respective energetic
and entropic functionals gives formally immediate access to the
thermodynamics for (in general) spatially inhomogeneous systems
governed by pairwise interparticle forces.

In conclusion, we have presented a formally exact variational
framework for thermal many-body physics. The potential energy
contributions due to the interparticle and external interactions are
treated on an equivalent footing. The existence and uniqueness of the
universal excess entropy functional $S_\rmexc[\rho,G]$, which carries
functional dependence on the density profile $\rho(\rv)$ and on the
global interparticle distance distribution $G(r)$, follows rigorously
from first principles. The theory retains formally all many-body
correlation effects that are generated by the Hamiltonian
\eqref{EQentropyHamiltonian} with pairwise interparticle potential
\eqref{EQentropyInterparticlePotentialPairwise} in equilibrium.

Future work could be addressed at designing functional machine
learning schemes to make the formal functional relationship accessible
in practice on the basis of simulation-based training.  Making
progress on the basis of analytical approximations is certainly a
highly challenging yet perhaps not hopeless task. The physics of
(quasi-)one-dimensional systems, where continuing progress is being
reported \cite{montero2019, montero2024, montero2024squarewellDisks,
  montero2024letter, montero2026dumbbell}, could provide important
guidance.  Note that the two arguably most successful free energy
density functional approximations describe either only entropy
(fundamental-measure theory for hard spheres) or only interparticle
potential energy (mean-field approximation). The present approach
offers a systematic framework to go beyond the simplistic {\it ad hoc}
sum of these two contributions.  {\ms Relating the present approach to
  Rosenfeld's concept of universality of the bridge functional
  \cite{rosenfeld1993bridge} can be important; see
  Refs.~\cite{simon2024patchy} for recent work and
  Ref.~\cite{schmidt2026entropyLong} for a brief overview.}

When applying the entropic framework to homogeneous bulk fluids,
significant simplifications arise due to $\rho(\rv)=\rm const$ and
$G(r)$ reducing to the (scaled) standard pair distribution function.
Percus' test particle trick~\cite{percus1962} allows one to access the
pair structure, via setting $V_\rmext(\rv)=\phi(|\rv|)$ in the
position Euler-Lagrange equation~\eqref{EQentropyEL2rhoG}. Entropic
meta-Ornstein-Zernike equations follow from functional differentiation
of the Euler-Lagrange equations and very simple excess entropy
approximations recover the standard second virial and mean-field
approximations, as presented elsewhere
\cite{schmidt2026entropyLong}. 
{\ms Also shown is the treatment of multi-component and of quantum
  systems, as well as the hyperfunctional theory for the entropy-based
  treatment of the thermal physics of general observables $\hat
  A$. Systems that interact via $n$-body potentials, which go beyond
  the pairwise contribution $\phi(r)$, are addressed via reduction
  operators $\hat G_n(R_n)$. These generalizations of the pairwise
  operator $\hat G(r)$ to the $n$-body case offer significant
  efficiency benefits via the reduced argument complexity of geometric
  parameters~$R_n$ over the explicit $n$-body density resolution. The
  excess entropy functional $S_\rmexc[\rho,\{G_n\}]$ acquires
  functional dependence on only those reduction fields $G_n(R_n)$ that
  are required to express the Hamiltonian, e.g., up to $n=3$ for the
  case of simultaneous presence of pair and triplet interparticle
  interactions. The theory remains closed on the functional level, as
  the equilibrium behaviour of the reduction fields is determined
  formally by extension of the pair distance Euler-Lagrange equation
  \eqref{EQentropyEL2rhoG} to the $n$-body case.

We summarize several salient points that are relevant for the future
use of the entropy functional approach.
i)~The unique splitting of the grand potential into clearly separated
entropy and energy contributions follows from first principles and it
remains structurally imprinted in all occurring direct and fluctuation
correlation functions. These objects offer a higher degree of
resolution than what is provided by conventional liquid state theory
\cite{hansen2013}.  This is important, as the entropy functional
theory retains the broad applicability of classical density functional
theory to general bulk and inhomogeneous equilibrium states.

ii)~While working with the excess entropy functional poses significant
theoretical challenges, given its universal applicability across all
forms of pair and external potentials, we emphasize that some of the
emerging complexity can be regarded as being offloaded onto the
formally exact mathematical structure. This is apparent in the
doubling of the Euler-Lagrange equations \eqref{EQentropyEL1rhoG} and
\eqref{EQentropyEL2rhoG} and in the quadrupling of Ornstein-Zernike
equations \cite{schmidt2026entropyLong} over the standard
framework. As emphasized throughout, the extensions include mere
scalar pair distance dependence, irrespective of the spatial
dimensionality.

iii)~Any practical application of the entropy functional theory
requires one to have a handle on the functional closure. Generalizing
the known exact one-dimensional free energy functionals
\cite{percus1976, robledo1981, vanderlick1989, percus1982, bakhti2012}
to the universal excess entropy functional constitutes clearly a very
significant theoretical challenge. We re-iterate that despite its
difficulty this task might not be hopeless, given the continuing
progress in exact solutions \cite{montero2019, montero2024,
  montero2024squarewellDisks, montero2024letter, montero2026dumbbell}.
Constructing approximative analytical forms can draw on the rich
heritage of classical density functional approximations, including the
local density and weighted density approximation, square-gradient
functionals, as well as fundamental-measure theories \cite{evans2016}.

iv) The arguably most direct route to practical applications could be
provided by machine learning, where classical density functional
theory proves to be a highly flexible, yet robust platform for
constructing training schemes that are both relatively simple and
powerful, such as the {\it local learning} method by Samm\"uller {\it
  et al.}~\cite{sammueller2023neural, sammueller2023whyNeural,
  sammueller2024pairmatching}.  Thereby the one-body direct
correlation functional $c_1(\rv;[\rho])$ is trained on the basis of
computer simulation results for randomized inhomogeneous systems. The
application within the present framework requires one to carry out
joint training of the two entropic direct correlation functionals
\eqref{EQentropyDirectEntropic1} and \eqref{EQentropyDirectEntropic2}
on the basis of results for the density profile $\rho(\rv)$ and the
pair distance histogram $G(r)$.  Utilizing a single neural network
with two outputs might be possible, following the lines of the recent
efficient learning of the bulk and interfacial physics in binary
Lennard-Jones mixtures \cite{robitschko2025mixShort,
  robitschko2026mixLong}. The detailed entropy functional structure
\cite{schmidt2026entropyLong} can then be put to use for the
systematic development of deep statistical mechanical testing,
regularization, and prediction schemes.

}

\bigskip

Bob Evans, Florian Samm\"uller, Stefanie M.\ Kampa, Johanna M\"uller,
Ana M. Montero, and Thomas Kriecherbauer are gratefully acknowledged
for useful discussions.  This work is supported by the DFG (Deutsche
Forschungsgemeinschaft) under Project No.~551294732.


\end{document}